\documentclass[letterpaper,journal]{IEEEtran}
\usepackage{amsmath,amssymb,amsfonts}
\usepackage{siunitx}
\usepackage{mathrsfs}
\usepackage{algorithmic}
\usepackage{algorithm}
\usepackage{array}
\usepackage[caption=false,font=normalsize,labelfont=sf,textfont=sf]{subfig}
\usepackage{textcomp}
\usepackage{stfloats}
\usepackage{url}
\usepackage{verbatim}
\usepackage{graphicx}
\usepackage{cite}
\usepackage{hyperref}
\begin{document}

\title{VoxelRF: Voxelized Radiance Field for Fast Wireless Channel Modeling}

% \author{Zihang Zeng,~Shu Sun,~\IEEEmembership{Senior Member,~IEEE},~and~Meixia Tao,~\IEEEmembership{Fellow,~IEEE}}

\author{Zihang Zeng,~Shu Sun,~\IEEEmembership{Senior Member,~IEEE},~Meixia Tao,~\IEEEmembership{Fellow,~IEEE},\\~Yin Xu,~\IEEEmembership{Senior Member,~IEEE},~and~Xianghao Yu,~\IEEEmembership{Senior Member,~IEEE}
        % <-this % stops a space
% \thanks{Received 25 February 2025; accepted 3 April 2025. This work was supported in part by the National Natural Science Foundation of China (NSFC)
% under Grant 62271310 and Grant 62431014; in part by the Science and
% Technology Commission Foundation of Shanghai under Grant 24DP1500702;
% and in part by NSFC under Grant 62001254. The associate editor coordinating
% the review of this article and approving it for publication was C.-W. Huang.
% \textit{(Corresponding authors: Shu Sun; Meixia Tao.)}}
% \thanks{This work is supported by the National Natural Science Foundation of China under Grants 62431014, 62271310, 62125108, and 62422111. (\textit{Corresponding authors: Shu Sun; Meixia Tao}).}
\thanks{Z. Zeng, S. Sun, M. Tao, and Y. Xu are with the School of Information Science and Electronic Engineering, Shanghai Jiao Tong University, Shanghai, 200240,
China  (emails: \{tarjan\_zeng, shusun, mxtao, xuyin\}@sjtu.edu.cn).}
\thanks{X. Yu is with the Department of Electrical Engineering, City University of Hong Kong, Hong Kong, China. (email: alex.yu@cityu.edu.hk).}
% \thanks{
% Ruifeng Gao is with the School of Transportation and Civil Engineering,
% Nantong University, Nantong 226019, China, and also with the Nantong Research Institute for Advanced Communication Technologies, Nantong
% 226019, China (e-mail: grf@ntu.edu.cn).}
% \thanks{Digital Object Identifier 10.1109/LWC.2025.3560446}
% <-this % stops a space
%\thanks{Manuscript received April 19, 2021; revised August 16, 2021.},
}

\maketitle
\begin{abstract}
Wireless channel modeling in complex environments is crucial for modern communication system design and deployment. Traditional channel modeling approaches face challenges in balancing accuracy, efficiency, and scalability, while recent neural approaches such as neural radiance field (NeRF) suffer from long training and slow inference. To tackle these challenges, we propose voxelized radiance field (VoxelRF), a novel neural representation for wireless channel modeling that enables fast and accurate synthesis of spatial spectra of received signals. VoxelRF replaces the costly multilayer perceptron (MLP) used in NeRF-based methods with trilinear interpolation of voxel grid-based representation and two shallow MLPs to model both propagation and transmitter-dependent effects. To further accelerate training and inference speed, we introduce an empty space skipping mechanism to reduce sampling in free space. Experimental results demonstrate that VoxelRF achieves competitive accuracy with significantly reduced computation and limited training data, making it more practical for real-time and resource-constrained wireless channel prediction.
\end{abstract}

% \vspace{-0.5cm}%%减小图片上间隔
\section{Introduction}
\IEEEPARstart{W}{ireless} channel modeling is an indispensable foundation for the design, optimization, evaluation, and deployment of communication systems. With the evolution of the sixth-generation (6G) wireless systems, spanning new spectrum bands and more complex propagation environments, advanced channel modeling approaches are required that possess high precision, high efficiency, and strong adaptability.

Conventional wireless channel modeling methods can be classified into three categories: deterministic, stochastic, and hybrid approaches.
The deterministic method models the wave propagation based on the electromagnetic (EM) theory \cite{sarkar2003survey}. For instance, ray-tracing (RT) can determine all possible routes between the transmitter (Tx) and receiver (Rx) based on the high-frequency approximation of Maxwell’s equations and computer-aided design models of the environment. However, such methods demand comprehensive knowledge of the environment’s geometry and the dielectric properties of materials, which can be challenging to obtain in practice. Moreover, RT often involves high computational costs and hence can hardly be applied in time-sensitive scenarios \cite{han2022terahertz}.
To address these challenges, the stochastic method divides propagation scenarios into different types and employs corresponding empirical models usually derived from measurements to capture channel characteristics. In addition, it has lower computational complexity, which allows fast channel model construction and system simulations. Nevertheless, it is achieved at the expense of accuracy and is difficult to provide site-specific channel characteristics.
% For example, statistical model cannot estimate angle of arrivals (AOAs). As a result, it cannot provide spatial spectra, which are heatmaps that characterize the energy distribution of the received signal from different AOAs.
Hybrid methods, which combine stochastic and deterministic approaches, are proposed to balance accuracy and complexity when dealing with complex environment. However, these methods still require geometry information of the environment.

Recently, the neural radiance field (NeRF) has emerged as a popular computer vision method for novel view synthesis, which uses limited images to learn a continuous optical field that renders novel views at arbitrary angles. 
% Remarkably, it learns a scene with a multilayer perceptron (MLP) to achieve implicit and continuous volumetric representation \cite{mildenhall2021nerf}. 
Specifically, for each scene, NeRF implicitly learns a continuous volumetric representation using a multilayer perceptron (MLP) to encode the spatial and appearance characteristics from different views.
Since light is a kind of EM wave, NeRF\textsuperscript{2}, proposed in \cite{zhao2023nerf2}, extends the radiance field concept to wireless channel, enabling the modeling of wireless radiance field to obtain spatial spectra  (power distribution of received signal at different angles) with a fixed Rx and a moving Tx. While NeRF\textsuperscript{2} demonstrates promising results, it inherits two drawbacks from NeRF: long training time and slow inference speed, due to the frequent query of a deep MLP for each sampling point along rays \cite{fridovich2022plenoxels}. Besides, the moving Tx leads to a varying wireless radiance field, which deviates from the nature of capturing static radiance field in NeRF \cite{mildenhall2021nerf}. Consequently, it requires significantly dense data for training.
NeWRF, proposed in \cite{lu2024newrf}, adopts the same method as NeRF\textsuperscript{2} but simplifies supervision by predicting wireless channel and limits ray directions to those derived from direction-of-arrival (DoA) measurements to support limited measurements-based reconstruction. However, it is still constrained by the vanilla NeRF model in terms of training and inference speed. The work \cite{li2024nera} proposes neural reflectance and attenuation fields (NeRA) to predict the radio map, which leverages the 3D environment information to skip sampling points in air and improve prediction speed. However, it requires a light detection and ranging scanner to acquire the point cloud of the environment.
WRF-GS+ \cite{wen2024wrf} employs 3D Gaussian splatting (3D-GS) \cite{kerbl20233d} to achieve real-time synthesis of spatial spectra. However, 3D-GS has limited capability in modeling high-frequency variations due to the inherently smooth and continuous nature of the Gaussian primitives and cannot cope with dynamic fields either.

To resolve the aforementioned issues, we propose VoxelRF, a novel voxelized radiance field framework for fast wireless channel modeling. The key innovations are as follows: First, we propose a dual-voxel representation for the wireless radiance field and utilize trilinear interpolation to replace the deep MLP query in NeRF-based approaches. Then, we design direction-aware signal mapping by utilizing two shallow MLPs to capture Tx-dependent effects and map the feature with the ray direction to the transmitted signal of each sample. Furthermore, we use empty space skipping to accelerate both training and inference speed.

% \begin{figure}[t]
% \centering
% \includegraphics[width=2.5in]{interaction.eps}
% \caption{Visualization of radio wave interactions between Tx and Rx in a 3D view.}
% \vspace{-0.5cm}%%减小图片上间隔
% \label{interaction}
% \end{figure}

% \vspace{-0.3cm}%%减小图片上间隔
\section{System Model}

% \begin{figure*}[!t]
% \centering
% \subfloat[]{\includegraphics[width=2.5in]{interaction}%
% \label{interaction}}
% \hfil
% \subfloat[]{\includegraphics[width=2.8in]{sampling}%
% \label{sampling}}
% \caption{Wireless channel model. (a) Radio wave interactions. (b) Ray sampling.}
% \label{system_model}
% \end{figure*}
We consider a scene in which an Rx is fixed at a certain location and a Tx is placed at different positions. For each position of the Tx, we can obtain a spatial spectrum at the Rx, which is equipped with a phased antenna array capable of forming a highly directional narrow beam and steering it to receive signals from a specific direction.
As EM waves propagate through the environment, they undergo various interactions such as reflection, diffraction, and scattering at the surfaces of objects. Therefore, the Rx receives signals from various directions. According to Huygens–Fresnel principle, these interaction points can be regarded as a new source of radiance. To compute the received signal at the Rx from a particular direction $\boldsymbol{\omega}_0=(\phi_0,\theta_0)$, where $\phi\in(0,\pi]$ and $\theta\in(0,\pi/2]$ correspond to the azimuthal and elevation angles, we perform sampling along this direction and treat each sample as a virtual Tx. The position of the $i$-th virtual Tx $\boldsymbol{x}_i$ is defined as:
\begin{equation}
\boldsymbol{x}_i = \boldsymbol{P}\!_{Rx} + r_i \boldsymbol{\omega}_0,
\label{x_i}
\end{equation}
where $\boldsymbol{P}\!_{Rx}$ denotes the position of the Rx, and $r_i$ is the distance between the Rx and the $i$-th sample.

To apply the concept of NeRF to wireless radiance field, we define an attenuation coefficient $\sigma(\boldsymbol{x}_i)$ at each sample $\boldsymbol{x}_i$, which represents the rate of loss of signal strength per unit distance due to dissipative effects. Additionally, we define the re-emitted signal $S(\boldsymbol{x}_i, -\boldsymbol{\omega}_0)$ as the signal radiated from point $\boldsymbol{x}_i$ along direction $-\boldsymbol{\omega}_0$. Therefore, the signal received from $\boldsymbol{x}_i$ can then be computed as follows:

\begin{align}
R(\boldsymbol{\omega}_0,\boldsymbol{x}_i) = T_i S(\boldsymbol{x}_i, -\boldsymbol{\omega}_0)
\label{R_omega_i}
\end{align}
where $T_i= \prod_{j=1}^{i-1} (\exp\left(-\sigma(\boldsymbol{x}_i) \delta_i\right))$ is the accumulated attenuation of signal from point $\boldsymbol{x}_i$ to the receiver, and $\delta_i$ is the distance between adjacent sample points. To obtain the total signal strength along $-\boldsymbol{\omega}_0$, the aggregated signal from $K$ sampling points is given by 
\begin{align}
R(\boldsymbol{\omega}_0) &= \sum_{i=1}^{K} R(\boldsymbol{\omega}_0,\boldsymbol{x}_i).
\label{R_omega}
\end{align}

% \begin{figure}[t]
% \centering
% \includegraphics[width=2in]{sampling}
% \caption{Spatial spectrum and ray sampling.}
% \label{sampling}
% \end{figure}

By aggregating the signals received from all possible directions $\boldsymbol{\Omega}$, we can compute the total received signal $R$ at the Rx as:
\begin{align}
R = \sum_{\boldsymbol{\omega} \in \boldsymbol{\Omega}} R(\boldsymbol{\omega})
= \sum_{\boldsymbol{\omega} \in \boldsymbol{\Omega}} \sum_{i=1}^{K} T_i S(\boldsymbol{x}_i, -\boldsymbol{\omega}_0).
\label{R}
\end{align}
While~\eqref{R_omega} provides the spatial spectrum at the receiver, which denotes the energy distribution of received signal power at the receiver, \eqref{R} computes the total received signal, from which the received signal strength indicator (RSSI) can be obtained. In this paper, we focus on the synthesis of the spatial spectrum for each Tx position. Therefore, the learning objective is to minimize the mean squared error (MSE) of the per-direction signal:
\begin{align}
\mathcal{L}_{\text {spectrum}}&=\frac{1}{|\boldsymbol{\Omega}|} \sum_{\boldsymbol{\omega} \in \boldsymbol{\Omega}}\|\hat{R}(\boldsymbol{\omega})-R(\boldsymbol{\omega})\|_{2}^{2},
\label{loss}
\end{align}
where $|\boldsymbol{\Omega}|=M\times N$ is the resolution of the spatial spectrum, corresponding to $M$ azimuth angles $\phi$ and $N$ elevation angles $\theta$, uniformly sampled over the unit upper hemisphere centered at the Rx. Therefore, the spatial spectrum with angular resolution of $1^\circ$ can be represented as a $360\times 90$ matrix $\mathbf{A}$, defined as
\begin{equation}
\mathbf{A}\!=\!\left[\begin{array}{@{}cccc@{}}
\mathbf{A}\left(0^{\circ}, 0^{\circ}\right) & \mathbf{A}\left(0^{\circ}, 1^{\circ}\right) & \cdots & \mathbf{A}\left(0^{\circ}, 89^{\circ}\right) \\
\mathbf{A}\left(1^{\circ}, 0^{\circ}\right) & \mathbf{A}\left(1^{\circ}, 1^{\circ}\right) & \cdots & \mathbf{A}\left(1^{\circ}, 89^{\circ}\right) \\
\vdots & \vdots & \ddots & \vdots \\
\mathbf{A}\left(359^{\circ}, 0^{\circ}\right) & \mathbf{A}\left(359^{\circ}, 1^{\circ}\right) & \cdots & \mathbf{A}\left(359^{\circ}, 89^{\circ}\right)
\end{array}\right],
\label{spatial spectrum}
\end{equation}
where $\mathbf{A}(\boldsymbol{\omega})=|R(\boldsymbol{\omega})|^2$.

% \vspace{-0.3cm}%%减小图片上间隔
\section{Proposed Algorithm}

% \begin{figure*}[!t]
% \centering
% \subfloat[]{\includegraphics[width=2in]{interp}%
% \label{interpolatiuon}}
% \hfil
% \subfloat[]{\includegraphics[width=4in]{network}%
% \label{network}}
% \caption{Overview of proposed voxel-based explicit–implicit hybrid algorithm. (a) Trilinear interpolation. (b) Deformation and radiance network.}
% \label{algorithm}
% \end{figure*}
In this section, we propose a voxel-based explicit–implicit hybrid representation for wireless signal radiance field, which enables significantly faster training and inference speed compared to prior NeRF-based approaches. Furthermore, we incorporate a deformation network to model the dynamic wireless radiance field caused by the moving Tx.
% \vspace{-0.3cm}%%减小图片上间隔
\subsection{Radiance Field Voxelization}
Instead of using a deep MLP to represent the radiance field, shown in Fig.~\ref{nerf2}, we employ two explicit voxel grids: an attenuation voxel grid $\boldsymbol{V}\!_{\mathrm{attn}} \in \mathbb{R}^{L_{x} \times L_{y} \times L_{z} \times 1}$ and a feature voxel grid $\boldsymbol{V}\!_{\mathrm{feat}} \in \mathbb{R}^{L_{x} \times L_{y} \times L_{z} \times F}$, where $(L_x, L_y, L_z)$ denotes the size of the voxel grid and $F$ is the dimension of the feature space. To guarantee the continuty of the radiance field, for each sample $\boldsymbol{x}_i$, we obtain its volume attenuation $\sigma(\boldsymbol{x}_i)$ and feature vector $\text{Feat}(\boldsymbol{x}_i)$ by trilinear interpolation over the neighboring eight voxels, shown in Fig.~\ref{voxelrf}:
\begin{align}
\sigma(\boldsymbol{x}_i)\! &=\! \operatorname{interp}(\boldsymbol{x}_i,\! \boldsymbol{V}\!_{\mathrm{attn}})\!:\! (\mathbb{R}^3, \mathbb{R}^{L_x \times L_y \times L_z \times 1})\! \rightarrow\! \mathbb{R}, \\
\text{Feat}(\boldsymbol{x}_i)\! &=\! \operatorname{interp}(\boldsymbol{x}_i,\! \boldsymbol{V}\!_{\mathrm{feat}})\!:\! (\mathbb{R}^3, \mathbb{R}^{L_x \times L_y \times L_z \times F})\! \rightarrow\! \mathbb{R}^F.
\label{interp}
\end{align}

Since trilinear interpolation is differentiable, gradients from the loss function are back-propagated through the interpolation weights to update the voxel values during the training stage.
Voxel grids are naturally suitable for representing sparse scenes, especially in practical wireless propagation environments, where most of the space is occupied by air that makes minimal contribution to the signal at the Rx. In comparison, the uniform sampling strategy used in NeRF\textsuperscript{2} makes a large number of sampling points fall into empty regions, leading to unnecessary computational overhead. To improve sampling efficiency, we adopt an empty space skipping strategy during both training and inference stages. Specifically, we skip the sample if the volume attenuation of this sample is below a predefined threshold $\tau$, which means the sample lies in empty space, i.e. air.

% \vspace{-0.5cm}%%减小图片上间隔
\subsection{Deformation Network}

% \begin{figure}[t]
% \centering
% \subfloat[\label{}]{\includegraphics[scale=0.27]{nerf2}}
% \subfloat[\label{}]{\includegraphics[scale=0.27]{voxelrf}}
% \caption{The performance of the block-wise linear property based combining vector on different numbers of blocks $Q$. (a) Aggregation MSE and average training symbols per parameter versus $Q$. (b) Test accuracy at round 200 versus $Q$.}
% \label{interpolatiuon}
% \end{figure}

\begin{figure*}[!t]
\centering
\subfloat[]{\includegraphics[scale=0.33]{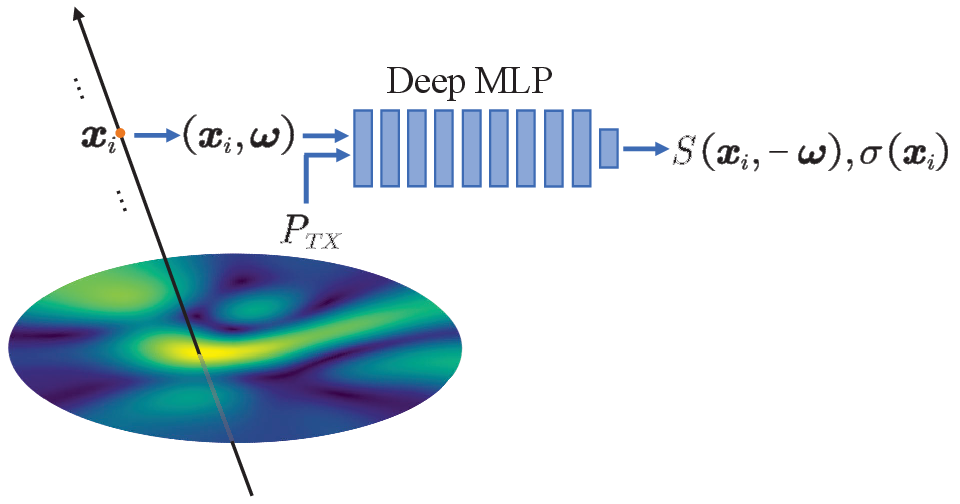}%
\label{nerf2}}
\hfil
\subfloat[]{\includegraphics[scale=0.33]{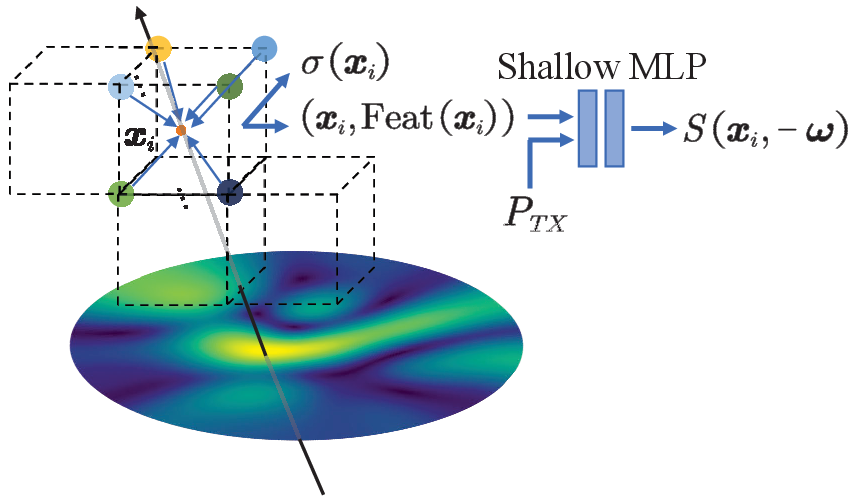}%
\label{voxelrf}}
\hfil
\subfloat[]{\includegraphics[scale=0.33]{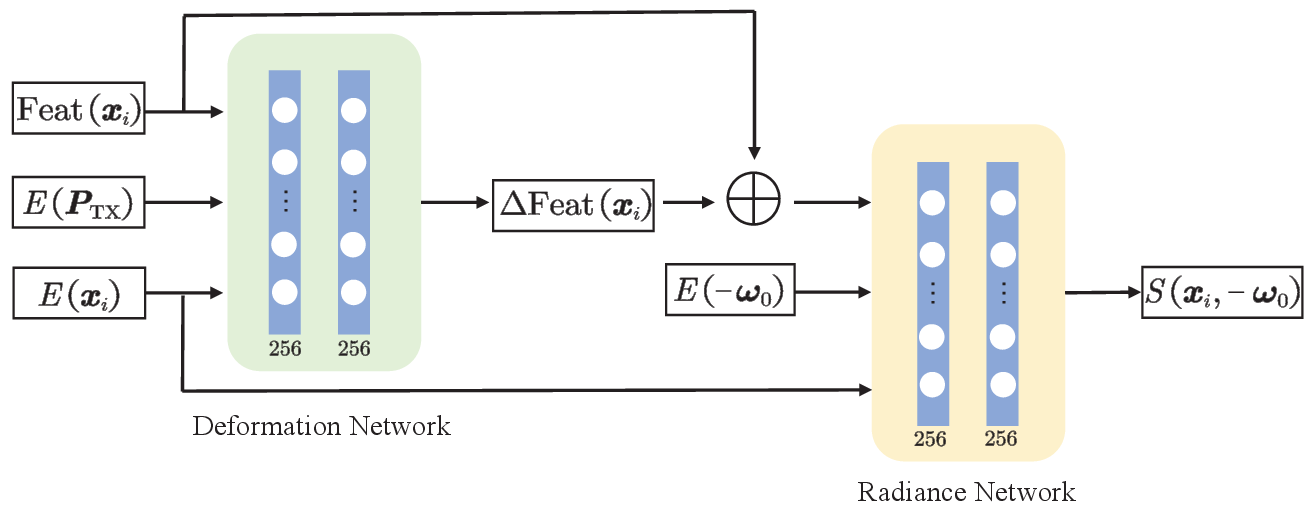}%
\label{network}}
\caption{Comparison between NeRF-based method and the proposed VoxelRF. (a) NeRF\textsuperscript{2}. (b) Proposed radiance field voxelization. (c) Proposed deformation network.}
\vspace{-0.5cm}%%减小图片上间隔
\label{interpolatiuon}
\end{figure*}

% \begin{figure}[t]
% \centering
% \includegraphics[width=1.5in]{interp}
% \caption{Trilinear interpolation.}
% \label{interpolatiuon}
% \end{figure}

As shown in Fig.~\ref{network}, we employ two shallow MLPs to predict the view-dependent signal emitted by each sample. The deformation network $D$ models the effect of Tx position on the local multipath characteristics, and outputs a deformation feature $\Delta\text{Feat}(\boldsymbol{x}_i)$. This allows the system to capture Tx position-dependent variations in the radiance field. In contrast, the original feature $\text{Feat}(\boldsymbol{x}_i)$ depends only on the spatial location $\boldsymbol{x}_i$, which encodes physical EM parameters such as conductivity and permittivity. The emitted signal $S(\boldsymbol{x}_i, -\boldsymbol{\omega}_0)$ is predicted by a radiance network $I$. Compared with the original NeRF-based framework, which directly inputs the Tx position into a radiance network, the introduction of the deformation network $D$ contributes to the generalization to Tx movements because it decouples the static and dynamic radiance field caused by moving Tx. It is worth noting that although the current framework does not consider the frequency-dependency of wave propagation, it can naturally incorporate frequency as an additional input feature to support multi-frequency modeling \cite{chen2025radio}.

% \begin{figure}[t]
% \centering
% \includegraphics[width=3in]{network}
% \caption{Deformation and radiance network.}
% \label{network}
% \end{figure}

We apply positional encoding to expand the dimensions of the Tx position $\boldsymbol{P}_{\text{Tx}}$, sample position $\boldsymbol{x}_i$, and view direction $-\boldsymbol{\omega}_0$, in order to capture high-frequency variations. The encoding is defined as:
\begin{align}
E(p) = \big( \sin(2^0 \pi p), \cos(2^0 \pi p), \cdots,\nonumber \\
        \sin(2^{L-1} \pi p), \cos(2^{L-1} \pi p) \big),
\end{align}
where $E(\cdot)$ is applied independently to each component $p$ of the input, and $L$ is a hyperparameter that controls the dimension of encoded input.

% \vspace{-0.3cm}%%减小图片上间隔
\subsection{Training Protocol}
\textbf{Progressive Learning}: A progressive learning strategy is adopted during training. Specifically, we begin with a coarse voxel grid containing $\left\lfloor L_{x} \times L_{y} \times L_{z} / 2^{M} \right\rfloor$ voxels, and progressively increase the number of voxels over $M$ stages. At each stage, the number of voxels is doubled, and both the attenuation grid $\boldsymbol{V}\!_{\mathrm{attn}}$ and the feature grid $\boldsymbol{V}\!_{\mathrm{feat}}$ are upsampled using trilinear interpolation until reaching the final resolution $(L_{x}, L_{y}, L_{z})$. This strategy reduces the risk of overfitting to the local optimum and improves training stability.

\textbf{Loss Function}: We add a background entropy loss $\mathcal{L}_{\text{bg}}$:
\begin{equation}
\mathcal{L}_{\text{bg}}\! =\! -\!\! \sum_{\boldsymbol{\omega} \in \mathcal{R}}\! \bigl( T_{K}(\boldsymbol{\omega}) \log T_{K}(\boldsymbol{\omega}) \!+\! \left(1 \!-\! T_{K}(\boldsymbol{\omega})\right) \log \left(1 \!-\! T_{K}(\boldsymbol{\omega})\right) \bigr),
\label{bg_loss}
\end{equation}
where $\mathcal{R}$ is the mini-batch of sampled rays of the spatial spectrum, $T_K(\boldsymbol{\omega})$ denotes the accumulated attenuation along ray $\boldsymbol{\omega}$. The background entropy loss encourages the network to make confident predictions about whether a ray is occluded, improving training convergence. Thus, the final loss function is expressed as:
\begin{equation}
\mathcal{L}= \mathcal{L}_{\text{spectrum}} + \lambda_{\text{bg}} \mathcal{L}_{\text{bg}},
\label{total_loss}
\end{equation}
where $\mathcal{L}_{\text {spectrum}}$ is given in \eqref{loss}, and $\lambda_{\text{bg}}$ is a weighting factor. 

% \vspace{-0.3cm}%%减小图片上间隔
\section{Experiments}
In this section, we implement and evaluate the proposed VoxelRF using real-world datasets.
% \vspace{-0.3cm}%%减小图片上间隔
\subsection{Implementation Details}
\subsubsection{Setup}
%模型设置、位置编码、训练设备、参数
The resolution of both the attenuation and feature voxel grids is set to $160\times 160\times 160$, with each voxel in the feature grid $\boldsymbol{V}\!_{\mathrm{feat}}$ having a feature dimension of $F = 24$.
Note that VoxelRF does not require any prior scene knowledge or dense sampling to build the voxel grids. Both the density and feature volumes are randomly initialized as learnable parameters and optimized directly from the measurement data.
Both the deformation network $D$ and the radiance network $I$ are implemented as two-layer MLPs, each with 256 channels. For activation functions, we apply the softplus function for attenuation $\sigma(\boldsymbol{x}_i)$ and the sigmoid function for emitted signal $S(\boldsymbol{x}_i, -\boldsymbol{\omega}_0)$. For positional encoding, we use $L=5$  for $\boldsymbol{P}_{\text{Tx}}$ and $\boldsymbol{x}_i$, and $L=4$ for $-\boldsymbol{\omega}_0$. The threshold for empty space skipping is set to $\tau = 1 \times 10^{-4}$, and the weighting factor for the background entropy loss is $\lambda_{\text{bg}} = 1 \times 10^{-4}$. Each ray is sampled at intervals equal to one-fourth of the voxel size.

We implement the proposed method in Python and train it on an NVIDIA GeForce RTX 3080Ti GPU. In addition, the Adam optimizer \cite{kingma2014adam} is used with a batch size of $1,024$ rays. The initial learning rate is set to $0.2$ for voxel grids and $2 \times 10^{-3}$ for the shallow MLPs, with exponential learning rate decay applied throughout the training. The training typically converges within $100k$ iterations.
\subsubsection{Datasets}
%RFID dataset
We use the open-source measured RFID dataset in NeRF\textsuperscript{2}, which captures wireless channel characteristics in an \SI{23}{\meter}~$\times$~\SI{24}{\meter}~$\times$~\SI{2}{\meter} indoor laboratory environment \cite{zhao2023nerf2}. A receiver is fixed and equipped with a $4\times 4$ antenna array operating at \SI{915}{MHz} and the Tx is an RFID tag, which is placed at random positions and continuously transmits RN16 messages. The dataset contains $6,123$ RFID positions, each with a $360\times 90$ spatial spectrum representing the received signal power distribution measured at the receiver. We use \SI{80}{\%} of the data for training and \SI{20}{\%} for testing, unless specified otherwise.
\subsubsection{Baselines}
%NeRF\textsuperscript{2}, WRF-GS+
We compare the proposed method with the following baselines:
\begin{itemize}
\item[$\bullet$] \textbf{NeRF\textsuperscript{2}}:
NeRF\textsuperscript{2} is the first method to use the neural radiance field for wireless signal modeling, which uniformly samples $64$ points in each ray and leverages a deep MLP in NeRF to map the sample position, Tx position, and view direction to volume attenuation and signal.
\item[$\bullet$] \textbf{WRF-GS+} \cite{wen2024wrf}:
WRF-GS+ is the latest method for spatial spectrum synthesis, which uses 3D-GS to reconstruct the wireless radiation field by employing 3D Gaussian primitives and neural networks to model the interactions between the environment and wireless signals.
\item[$\bullet$] \textbf{Deep Convolutional Generative Adversarial Network (DCGAN)} \cite{radford2015unsupervised}:
DCGAN is a widely used generative model consisting of a generator and a discriminator. The generator produces spatial spectrum maps based on the transmitter position, while the discriminator evaluates whether the spectrum comes from real measurements or from the generator.
\end{itemize}

% \vspace{-0.3cm}%%减小图片上间隔
\subsection{Evaluation}
\begin{figure}[!t]
\centering
\includegraphics[width=3.4in]{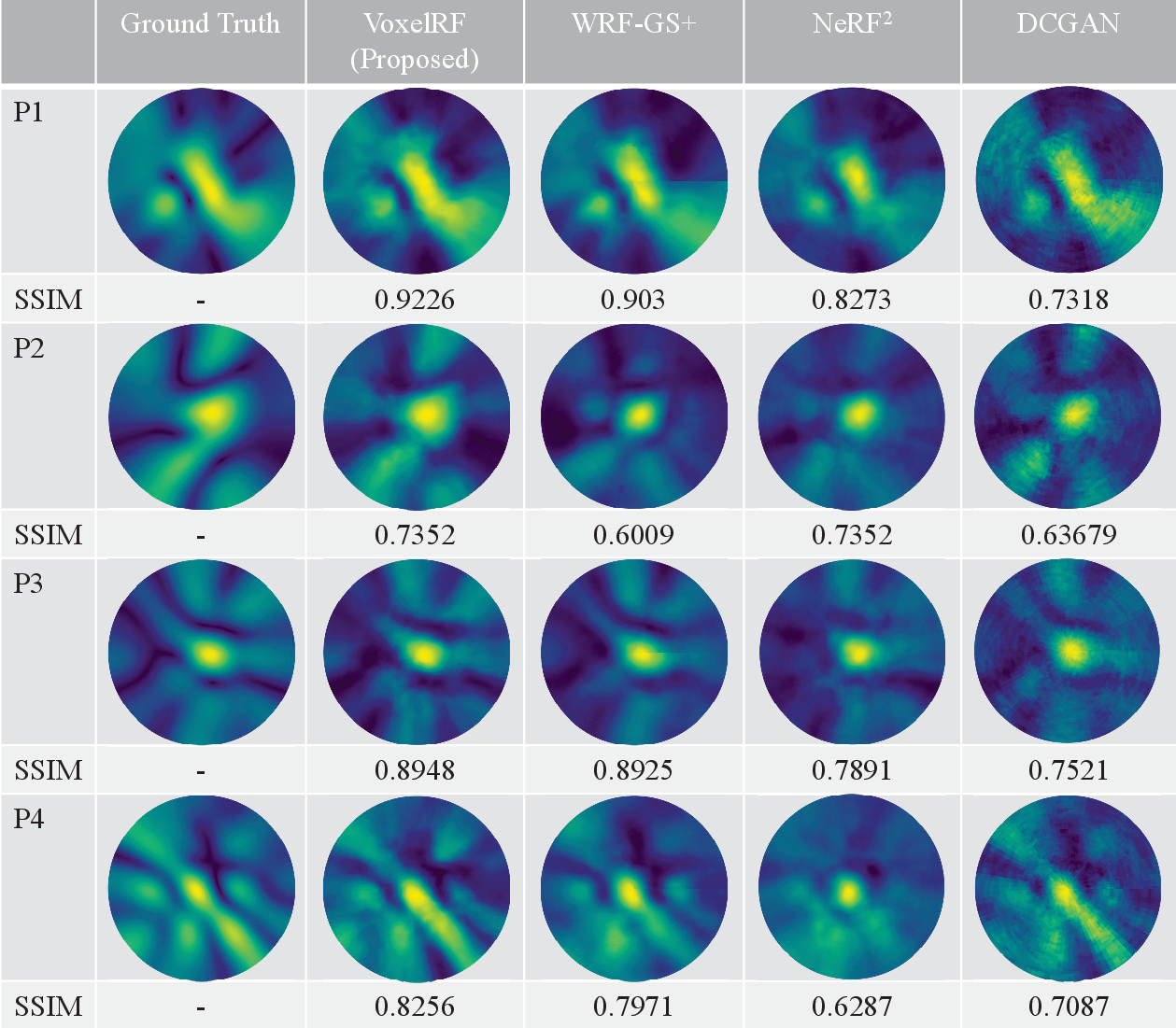}
% \vspace{-0.3cm}%%减小图片上间隔
\caption{Comparison of spatial spectra at four Tx positions.}
\vspace{-0.3cm}%%减小图片上间隔
\label{spectrum}
\end{figure}

First, we compare the spatial spectra generated by VoxelRF, WRF-GS+, NeRF\textsuperscript{2} and DCGAN, as shown in Fig.~\ref{spectrum}, to provide an intuitive appraisal. It is obvious that the spectra generated by VoxelRF are the closest to the ground truth, which exhibits sharper transitions between high and low intensity regions, indicating better preservation of high-frequency components. This suggests that VoxelRF is capable of capturing the rapid spatial fluctuations from small-scale fading effects more accurately. In contrast, although WRF-GS+ achieves better performance than NeRF\textsuperscript{2}, it produces blurry edges due to the utilization of Gaussian primitives, which are inherently smooth and continuous, leading to a loss of sharp detail, especially at boundaries.

\begin{figure}[t]
\centering
\includegraphics[width=3.4in]{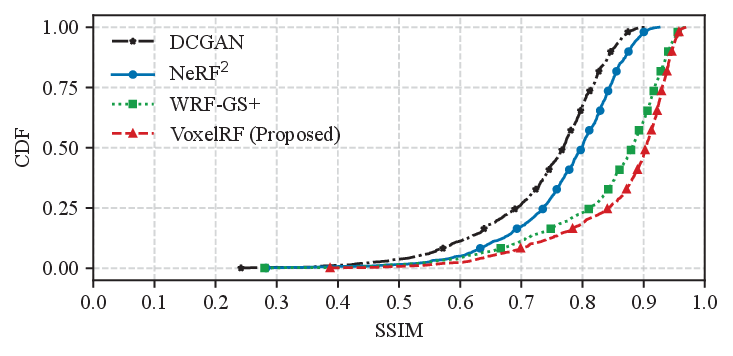}
\vspace{-0.3cm}%%减小图片上间隔
\caption{SSIM distribution of spatial spectrum with different methods.}
\vspace{-0.3cm}%%减小图片上间隔
\label{ssim}
\end{figure}

Then, we evaluate the performance of different methods using the structural similarity index measure (SSIM). SSIM is an image quality metric used to assess the similarity between two images based on three fundamental components: luminance, contrast, and structure \cite{wang2004image}. Fig.~\ref{ssim} illustrates the cumulative distribution function (CDF) of the SSIM between the synthesised spatial spectrum and the ground truth. The median SSIM of DCGAN, NeRF\textsuperscript{2}, WRF-GS+ and VoxelRF is $0.7691$, $0.7991$, $0.8813$ and $0.9035$, respectively. VoxelRF achieves the best performance because it employs the deformation network to separate static and dynamic field caused by the moving Tx. In contrast, NeRF\textsuperscript{2} is designed for static scenes, which means it possesses limited capabilities in capturing the influence of Tx position variations.

\begin{figure}[t]
\centering
\includegraphics[width=3.4in]{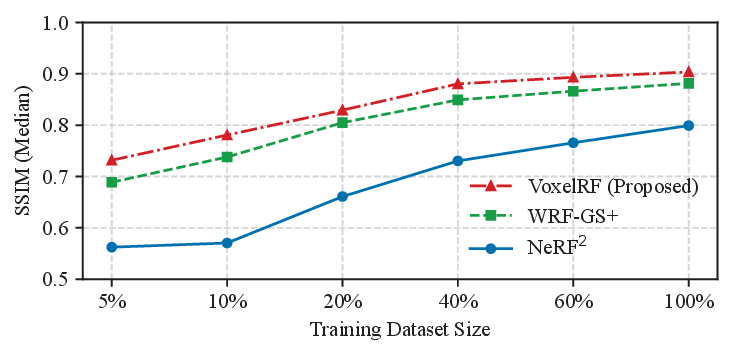}
\vspace{-0.3cm}%%减小图片上间隔
\caption{Median values of SSIM under different training dataset sizes.}
\vspace{-0.3cm}%%减小图片上间隔
\label{ssim_vs_datasize}
\end{figure}

Next, we compare the median SSIM values across different training dataset sizes, as shown in Fig.~\ref{ssim_vs_datasize}. It is evident that VoxelRF outperforms the baseline methods for each size of the training dataset, which shows the robustness of VoxelRF in limited-data scenarios. In addition, it can be seen that the proposed model converges with about $60\%$ of the training data, indicating its data efficiency. This is because compared to the pure implicit representation in NeRF\textsuperscript{2}, VoxelRF employs explicit voxel grids to represent the radiance field and interpolation to ensure local continuity, which regularizes the learning process and prevents the model from overfitting when the training data is limited.

Finally, we compare the training and inference speed of the spatial spectrum in Table~\ref{time}, which shows significant advantages of the proposed method. Specifically, VoxelRF reduces the training time to only $20$ minutes, which is $4\times$ faster than DCGAN, $8\times$ faster than WRF-GS+ and nearly $47\times$ faster than NeRF\textsuperscript{2}. In terms of inference, it achieves an inference time of $0.036$ seconds, which is $10\times$ faster than NeRF\textsuperscript{2}, nearly $3\times$ faster than DCGAN, and is comparable to WRF-GS+. This is because VoxelRF replaces the deep MLP in NeRF\textsuperscript{2} with trilinear interpolation of the voxel grid and skips the sample points with low volume attenuation. These results highlight VoxelRF’s potential for deployment in time-sensitive or resource-constrained applications.
In terms of the attenuation threshold $\tau$, we observed that the model runs out of memory (OOM) if $\tau$ is set to 0, and varying $\tau$ had little effect (less than $0.1\%$ variation) on the final performance while providing consistent speedup during both training and inference stages, which indicates that the model only skips regions with negligible contribution and does not remove meaningful multipath components.

\begin{figure}[!t]
\centering
\includegraphics[width=3.4in]{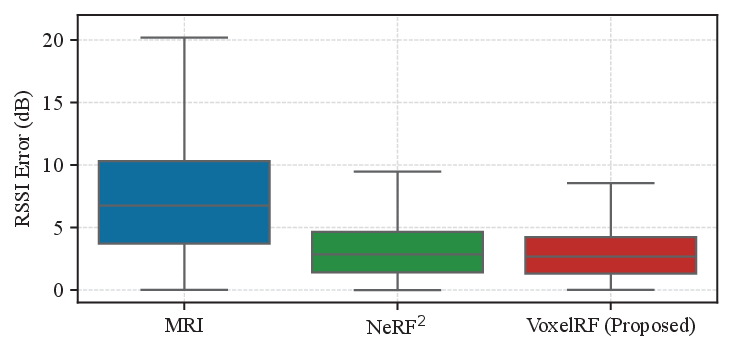}
\vspace{-0.3cm}%%减小图片上间隔
\caption{RSSI prediction error of MRI, NeRF\textsuperscript{2}, and VoxelRF.}
\vspace{-0.5cm}%%减小图片上间隔
\label{rssi}
\end{figure}

\begin{table}
\begin{center}
\caption{The training and inference time of different methods.}
% \vspace{-0.1cm}%%减小图片上间隔
\label{time}
{
\begin{tabular}{| c | c | c | c |}
\hline
\textbf{Method} & \textbf{SSIM} & \textbf{Training Time} & \textbf{Inference Time} \\
\hline
NeRF\textsuperscript{2} & 0.7991  & 15h30m & 0.39s \\
\hline
WRF-GS+ & 0.8813 & 2h40m  & 0.0085s \\
\hline
DCGAN & 0.7691 & 1h20m  & 0.09s \\
\hline
VoxelRF ($\tau$=0) & \text{OOM}  & -  & - \\
\hline
VoxelRF ($\tau$=10\textsuperscript{-5}) & 0.9025 & 20m21s  & 0.038s \\
\hline
VoxelRF ($\tau$=10\textsuperscript{-4}) & 0.9035 & 20m01s  & 0.036s \\
\hline
VoxelRF ($\tau$=10\textsuperscript{-3}) & 0.9029 & 19m19s  & 0.033s \\
\hline
\end{tabular}
}
\vspace{-0.5cm}%%减小图片上间隔
\end{center}
\end{table}

% \vspace{-0.4cm}%%减小图片上间隔
\subsection{Case Study: RSSI Prediction}
Since spatial spectrum denotes the power distributed in all directions, we can obtain the received signal strength by aggregating the spatial spectrum via \eqref{R}. We test the performance of VoxelRF using the measured bluetooth low energy (BLE) dataset in NeRF\textsuperscript{2}. In the BLE dataset, $21$ RXs and a moving Tx are placed in a 15,000 ft\textsuperscript{2} indoor facility. The Rx records the measured RSSI as Tx moves around \cite{zhao2023nerf2}. The dataset consists of $2,000$ valid measurements (RSSI above \SI{-100}{dBm}), $1,600$ for training and $400$ for testing. The baselines in this study are NeRF\textsuperscript{2} and model-based radio interpolation (MRI), which uses a radio propagation-based learning model to interpolate RSSI values at unvisited locations \cite{shin2014mri}. Fig.~\ref{rssi} shows the RSSI prediction error of VoxelRF, NeRF\textsuperscript{2} and MRI on the BLE dataset, where the RSSI error is the absolute difference between the predicted and the true RSSI value. Specifically, the median RSSI error of VoxelRF is $\SI{2.66}{dB}$ (25\textsuperscript{th} percentile: $\SI{1.31}{dB}$, 75\textsuperscript{th} percentile: $\SI{4.22}{dB}$), while the median errors of NeRF\textsuperscript{2} and MRI are $\SI{2.85}{dB}$ (25\textsuperscript{th} percentile: $\SI{1.41}{dB}$, 75\textsuperscript{th} percentile: $\SI{4.64}{dB}$) and $\SI{6.74}{dB}$ (25\textsuperscript{th} percentile: $\SI{3.72}{dB}$, 75\textsuperscript{th} percentile: $\SI{10.31}{dB}$), respectively. The superior performance of VoxelRF in RSSI prediction highlights its strong potential for the radio map construction task.

% \vspace{-0.3cm}%%减小图片上间隔
\section{Conclusion}
In this paper, we introduce VoxelRF, a fast and efficient wireless channel modeling approach that can accurately predict the received signal spatial spectrum and RSSI with a moving Tx. VoxelRF achieves state-of-the-art performance on both RFID and BLE datasets, with the shortest training time and competitive inference speed, by voxelizing the radiance field and decoupling static environmental features from dynamic Tx effects via a deformation network.

Although the proposed VoxelRF shows significant potential for wireless channel modeling, two key challenges remain. First, similar to other NeRF–based methods, VoxelRF is optimized per scence and therefore exhibits limited cross-scene generalization. Second, the current framework does not incorporate phase information, making it unable to capture coherent interference effects of electromagnetic fields.

% \vspace{-0.3cm}%%减小图片上间隔
\bibliographystyle{ieeetr}
\bibliography{lookup}

\vfill

\end{document}